# Title: Real-space pairing through a confined local nematic state in cuprate superconductors

**Authors:** Huazhou Li[1†], Han Li[1†], Siyuan Wan[1], Zhaohui Wang[1], Huan Yang[1*], Hai-Hu Wen[1*]

**Affiliations:**

[1]National Laboratory of Solid State Microstructures and Department of Physics, Collaborative Innovation Center of Advanced Microstructures, Nanjing University, Nanjing 210093, China

*Corresponding author. Email: huanyang@nju.edu.cn (H. Y.); hhwen@nju.edu.cn (H.-H.W.)

[†]These authors contributed equally to this work.

**Abstract:** The pairing mechanism of high−temperature superconductivity in cuprates is regarded as one of the most challenging issues that we are facing now. The core issue is about how the Cooper pairs are formed. Here we report the spin-resolved tunneling measurements on extremely underdoped $Bi_2Sr_{2-x}La_xCuO_{6+\delta}$. Our data reveal that, when holes are doped into the system, the antiferromagnetic order is destroyed together with the filling up of the charge transfer gap by some density of states (DOS) with a maximum at about 200 meV. Consequently, an electronic structure with $4a_0 \times 4a_0$ basic plaquettes gradually emerges, and they become more populated versus further hole doping. In the central part of each plaquette, there are three unidirectional bars (along the Cu-O bond) enlightened by the enhanced peaks of coherence-like DOS at about ±25 meV, and the intensity is especially pronounced at oxygen sites. We argue that this confined nematic state within one $4a_0 \times 4a_0$ plaquette constitutes naturally the local pairing format. Our work shed new light in unraveling the longstanding puzzle of pairing mechanism in cuprate superconductors.

**One-Sentence Summary:** A confined nematic state is observed as an evidence of the local pairing in extremely underdoped cuprates.

According to the Bardeen-Cooper-Schrieffer (BCS) theory, superconductivity is induced by the condensation of Cooper pairs. In conventional superconductors, the Cooper pairing is established through exchanging phonons between two electrons with opposite momentum, which has an upper limit of critical temperature ($T_c$) at ambient pressure. The discovery of high-$T_c$ cuprate superconductors breaks the limit of $T_c$, and challenges the picture of phonon-mediated pairing. Afterwards, to understand the paring mechanism in cuprates becomes one of the most interesting and challenging issues in condensed matter physics. Several theoretical models have been proposed in order to tackle this problem. Based on the Hubbard model, the earliest scenario concerning the pairing is the resonating-valence-bond (RVB) model in which the simultaneous spin-singlet pairing is formed by the antiferromagnetic (AF) superexchange interaction of local spins of Cu ions (*1,2*). Other magnetic pairing pictures are either based on *t-J* model (*3*) or concern the exchange of AF spin fluctuations (*4-6*). In the strong-coupling limit, the pairing could be very local, thus the local pairing picture provides a great potential to explain the high-$T_c$ superconductivity. Of particular interest are the spatial inhomogeneity and the phase separation of electronic states, which can be induced by the frustrated quantum interactions in cuprates (*7,8*), and can even lead to some forms of the local pairing in the hole-doped but non-superconducting cuprates. A picture concerning pair-density-wave (PDW) of *d*-wave Cooper pairs without global phase coherence was proposed (*9*) in order to explain the real-space electronic modulations in the pseudogap regime (*10*) observed by scanning tunneling microscopy/spectroscopy (STM/S). Later, the PDW phase was discovered in the superconducting state of cuprates (*11,12*) , and it shows a new form of the possible local pairing with an entirely different form of broken symmetry from the previous theoretical proposal (*9*). In cuprates, although it is widely accepted that the high-$T_c$ superconductivity is induced by doping holes or electrons to the AF Mott insulator, it remains however unclear (i) how is the Mottness related gap filled together with the vanishing of the long range AF order? and (ii) whether the local Cooper pairing really exists and in what form?

Thanks to the spin-resolving capability of the spin-polarized (SP) STM/STS developed recently in our group (*13,14*), we report here a close relationship between the AF order and the Mottness derivative gap; meanwhile we show the emergence of a precursor superconductivity-related gap within a $4a_0 \times 4a_0$ basic plaquette with an internal nematic state. We argue that this may constitute the basic form of the local pairing.

The $Bi_2Sr_{2-x}La_xCuO_{6+\delta}$ (La-Bi2201) samples were grown with a traveling-solvent-floating-zone technique (*15*). In our present studies, we use the samples with nominal compositions of $x = 1$ and 0.85 which correspond to hole doping levels of $p = 0.08$ and 0.10, respectively. For simplicity, we denote these samples as UD08 and UD10. Because of slight oxygen deficiency which leads to extra underdoping, the real doping levels in our studied samples are slightly lower than $p = 0.08$ and 0.10, that is why even for UD10 we do not see a superconducting transition yet, but it is really in the vicinity of superconductivity (*15,16,17*). The phase diagram of La-Bi2201 is shown in Fig. 1A, it is clear that the AF Mott insulating phase is almost vanished and the superconducting phase is just about to appear at the doping level of $p = 0.10$ (*15,16,17*).

The temperature dependent resistivity of these two samples shows clear insulating features (Fig. 1B), and UD08 exhibits a stronger insulating behavior than that of UD10. We use the SP-STM to investigate the topographic images of these two kinds of samples (fig. S1) by using Cr tips polarized with magnetic fields along opposite directions (*13,14*). After cleaving the sample in ultrahigh vacuum, the exposed surface is the Bi-O layer, and the Bi atoms are assumed to correspond to the positions of Cu atoms in the Cu-O layer underneath (*18-21*). Thus any magnetic moment detected by SP-STM should arise from the Cu-O layer. The spin-difference topographies of the samples are shown in Fig. 1, C and D. The AF order can be clearly seen in some areas of UD08 (Fig. 1C), but it is almost invisible in UD10 (Fig. 1D). Such difference is clearer in the Fourier transformed (FT) patterns of the spin-difference topographies (insets of Fig. 1, C and D), i.e., the AF order is clear and commensurate with the period of $\sqrt{2}a_0$ along the diagonal direction of the lattice in UD08, but it is almost absent in UD10.

Next we investigate how does the AF order correlate with the Mottness derivative gap through measuring the local spectrum. The AF order can be observed in some areas in UD08. Figure 2A shows the spin-difference topography measured in another area of UD08. The AF order is weak in the region surrounded by the dashed line. The differential conductance (d$I$/d$V$) mapping has been measured at 100 mV in the same region (Fig. 2B), and the local intensity in the mapping is roughly proportional to the local density of states (LDOS). One can see that the intensity of LDOS at 100 meV gets enhanced in the region with vanished AF order. In contrast, the LDOS is almost zero in the region with the AF order. In addition, in the region without the AF order, some patterns with dimensions of $4a_0 \times 4a_0$ emerge, and this newly emergent patterns are clearer in the inverse Fourier transform image (Fig. 2C, and fig. S2C) or measured with a low energy. The $4a_0 \times 4a_0$ periodic patterns have been observed as checkerboard electronic modulations in underdoped superconducting cuprates (*22-25*), and they are also observed in the vortex cores in optimal doped cuprate Bi$_2$Sr$_2$CaCu$_2$O$_{8+\delta}$ (*26*). Therefore, the checkerboard electronic modulations may be closely related to superconductivity. It should be noted that our samples are highly insulating, thus the emergence of $4a_0 \times 4a_0$ plaquettes excludes the possibility that these plaquettes originate from the nesting effect of Fermi surface (*24*). Figure 2C shows a set of tunneling spectra measured along the arrowed line in Fig. 2A. In the region with an obvious AF order, the spectra are featureless showing a background within the Mottness derivative gap in a wide voltage range (up to about 1.5~2.0 V). In cuprates, it is known that the oxygen *p*-orbital hybridizes with the lower Hubbard band of Cu $3d_{x^2-y^2}$ orbital, leading to a charge transfer gap $\Delta$ in the scale of 1.5~2.0 V between the charge transfer band (CTB) and the upper Hubbard band (UHB) (*27*). The gap obtained from our experimental data in UD08 is consistent with the theoretically expected value. When the tip moves to the region with vanished AF order and presence of the checkerboard modulations, one can see that the gapped energy range shrinks. In addition, the LDOS increases from both sides of Fermi level, and a hump-like feature of density of states (DOS) appears on the spectra at about 150-200 mV. This is consistent with previous observations in the similar system in very underdoped region (*28*). In order to illustrate this emergent in-gap states more clearly, we take the first spectrum in the AF region as the background and subtract it from all other spectra, and show the data in Fig. 2D. One can see that the enhancement of DOS near the Fermi level is clearer in the narrow voltage range, showing a feature of large pseudogap. Such features of suppressing the charge transfer gap and enhancing DOS near the Fermi level in underdoped samples have also been found by another group (*28*). Here, for the first time, we illustrate that these features are closely related to the vanishing of the AF order, and this indeed verifies that the parent phase of cuprates is an antiferromagnetic Mott insulator.

In the sample with a slightly higher doping level of $p = 0.10$, the AF order almost disappears and more $4a_0 \times 4a_0$ plaquettes appear in the mapping of LDOS with a bias of 25 mV (Fig. 3A). Checkerboard modulations appear in about half area of the presented region, and the other half area is featureless with weak intensity of LDOS at this energy. Upon closer inspection, each $4a_0 \times 4a_0$ plaquette exhibits a very unique pattern, i.e., there are two or three unidirectional bars of enhanced DOS within the plaquette forming a nematic electronic state. The distance between two neighbored bars is about $4a_0/3$, which leads to an incommensurate modulation. Here in each plaquette, three bars are all along one of the crystalline axes (*a*- or *b*-aixs, namely the Cu-O-Cu direction), hence, the existence of such unidirectional bars breaks the fourfold symmetry of the crystal. However, these nematic patterns are orthogonal to each other among different plaquettes. Figure 3B shows the FT pattern of the d*I*/d*V* mapping in Fig. 3A. The spots of $Q' = (\pm 1/4, 0)2\pi/a_0$ or $(0, \pm 1/4)2\pi/a_0$ correspond to the $4a_0 \times 4a_0$ checkerboard modulations; those of $Q'' = (\pm 3/4, 0)2\pi/a_0$ or $(0, \pm 3/4)2\pi/a_0$ correspond to the internal $4a_0/3$ nematic modulations. In Fig. 3C, we show a set of spectra along the arrowed line across one $4a_0 \times 4a_0$ plaquette as marked in Fig. 3A. In the featureless region, the spectra show Mottness gapped feature which is similar to the AF region of UD08. In the region with the enlightened DOS, namely the plaquette, wide spreading DOS with a maximum at about 200 mV appear. This forms a pseudogap feature. In Fig. 3D, we show the same line-scan spectra by subtracting the one in the featureless region (Mottness gapped region). One can see that a hump-like feature of DOS appears near the Fermi level, which indicates the enhancement of DOS near the Fermi level in the $4a_0 \times 4a_0$ plaquette. This type of nematic internal structure was also observed in overdoped Bi2201 (*29*); the authors there regarded this kind of pattern as stripy structure and linked them with the pair density wave. In addition, the observed nematic structure here is local and confined within one plaquette, which seems different from the "stripe phase" seen by neutron scattering in $La_{1.6-x}Nd_{0.4}Sr_xCuO_4$ (*30*). In the "stripe phase", there are hole derivative charge chains neighbored by AF patterns or stripes with anti-phase spin directions on the boundary. This has not been seen in our spin-deference topographies. However, our observation about the internal nematic states is in general consistent with the charged magnetic domain or stripy patterns of electronic states in lightly underdoped Mott insulator (*31,32*). Based on the descriptions above, we can outline a general evolution of the electronic states versus hole doping. There are generally three types of electronic states on the surface: (i) the long range AF ordered regions with charge transfer gapped feature; (ii) the regions without AF order but with finite DOS in a wide energy region (a hump appears in the energy scale of ~200 meV); (iii) the regions with the $4a_0 \times 4a_0$ plaquettes and internal nematic electronic structures appearing especially at low energies.

In the following, let's have a close scrutinizing on the internal electronic structure of a $4a_0 \times 4a_0$ plaquette measured by a tungsten tip. Figure 4A shows the d*I*/d*V* mapping measured at 25 mV in an area of UD10, one can see many such plaquettes, within some plaquettes we can clearly see the unidirectional bars of DOS. The black dots here denote the positions of Bi atoms measured from the topography (fig. S3), and as aforementioned, they are supposed to correspond to the positions of Cu atoms in the Cu-O layer underneath the Bi-O layer (*18-21*). The oxygen sites on the Cu-O layer are taken in the middle of two neighboring Cu sites. One can see that there are three enlightened bars of DOS with the distance of an incommensurate period of about $4a_0/3$, and the central bar locates on and aligns along the central Cu-O-Cu chain. We select one of these plaquettes shown in Fig. 4A to do more detailed measurements of spectra, as shown in Fig. 4B, the three bars of higher DOS within the plaquette can be clearly seen. Then spectra are measured along different Cu-O-Cu chains numbered as b1 to b5 and shown in Fig. 4, C to G. The spectra measured in the $4a_0 \times 4a_0$ plaquette generally show a finite DOS with a V-shaped feature in the displayed energy window spanning to 240 meV. When the spectra are measured along b1 and b5, namely on the Cu-O-Cu chains at the edge of the plaquette, one can see that it reveals a large V shape of LDOS with a maximum in the region of about 200 mV (Fig. 4, C and G). When we move to b2 and b4, a pair of low-energy coherence-like peaks of d$I$/d$V$ arise in the central region with the peak energy at about 25 meV (Fig. 4, D and F). The most pronounced coherence-like peaks appear on b3 (Fig. 4E), and that is on the central bar (also the central Cu-O-Cu chain). This is one of the key findings here. A comparison of the spectra measured on one of the central oxygen (O2) and the copper at the edge (Cu5) along b3 is shown in Fig. 4H, and a drastic difference can be clearly seen. To emphasize the coherence-like peaks, we take a subtraction with the spectrum at Cu5 from that at O2, and show it as the inset in Fig. 4H. A sharp coherence peak at about 25 mV can again be clearly seen. On the negative bias side, this coherence peak is also present, but it looks less pronounced compared with the positive side, which is induced by a much steeper background on the side of negative bias due to the charge transfer band. The systematic evolution of spectra within the selected plaquette shown in Fig. 4B gives just a representative one, and control experiments have been repeated in more than 10 plaquettes. It clearly demonstrates that there is an internal nematic structure with enlightened bars of LDOS along the Cu-O-Cu chains, and the central one has the highest coherence-like peaks at about 25 mV.

We have witnessed that the coherence-like peak at about 25 mV is the highest at the oxygen O2 (Fig. 4E). In order to figure out the spatial dependence of LDOS more rigorously, we digitize the spatially dependent differential conductance at 25 mV along b3 in the central bar of the selected plaquette and show it in Fig. 5A. Surprisingly, some peaks and knees of differential conductance measured at 25 mV appear at or near the oxygen sites (O1, O2 and O3). This is another key finding in our present study, strongly suggesting that the emerging coherent weight through hole doping arises mainly at the oxygen sites. We then use the multi-peak Gaussian fitting method to obtain the possible peak positions of the experimental data (red circles in Fig. 5A), and the fitting result is shown as the blue solid line. One can see that the fitting curve coincides with the experimental results very well, and the resultant three Gaussian functions have the central positions very near the three oxygen atoms. We have also conducted line cuts along the central bars of different plaquettes, most of them show the peaks or knees of differential conductance on the oxygen sites (eight of them are shown in fig. S4 as control experiments). It should be noted that the experimental data obtained by using a SP Cr tip have the similar conclusion (fig. S5). One may argue that, the Cu atoms may not sit exactly below the Bi atoms, thus the measured differential conductance at the Bi and O atoms in the top layer may not reflect the signal from the Cu and O atoms underneath. To remove this concern, we take the data of spatial dependence of differential conductance along a straight line in the middle of the central bar which is supposed to coincide with a perfect Cu-O-Cu chain (fig. S6). One can see that the peaks or knees of d$I$/d$V$ still locate near oxygen sites, just like that shown in Fig. 5A. Our data directly prove that the doped holes are really going into the oxygen orbitals, as pointed out by Emery (*33*) and Zhang-Rice (*34*). Thus we can conclude that the doped holes induce some coherent-like weight of charge freedom mainly accommodating on oxygen sites.

In our experiments, we have illustrated three key observations. Firstly, we find that the AF order is closely related to the Mottness derivative charge transfer gap; when the AF order is diminished by hole doping, this gap will be filled up by DOS in a wide range of energy. Secondly, at a higher doping level, for example in UD10, the AF order is destroyed, the charge transfer gap is completely filled up by finite DOS with a wide spread spectrum and a maximum occurring at around 200 mV, we call this as a large pseudogap; meanwhile, some $4a_0 \times 4a_0$ plaquettes of DOS are formed; within each plaquette, an internal nematic electronic state is observed with coherence-like peaks of DOS at energies of about 25 meV. It is interesting to note that, this large pseudogap feature was also observed in the iridate system $(Sr_{1-x}La_x)_2IrO_4$ when the Mott gap is filled by doping electrons (*35*), showing a universal behavior in doping a Mott insulator. Thirdly, the highest coherence-like peaks appear on the oxygen sites, implying that the holes are doped into the oxygen orbitals. Since this kind of differential conductance peaks look like the coherence peaks in the superconducting samples, we strongly argue that this may cast as the precursor gap for superconductivity, such as the local pairing gap. Because the superfluid is still very low in present samples, thus the system cannot enter a superconducting condensation state, and we do not see a superconducting transition yet. In this context, the local and confined nematic electronic states observed here may share the same origin as the PDW observed in superconducting state (*11,12,29*). Based on these observations, a schematic plot of the local DOS is shown in Fig. 5B with the yellow-green elongated ellipses highlighting the locations where higher deferential conductance is observed. This unique nematic state within a $4a_0 \times 4a_0$ plaquette may reflect a local pairing format of doped holes. We argue in this way because of the following reasons. (1) Superconductivity is just about to appear when the doping level is about $p = 0.10$. By counting the hole densities, the state with $p = 0.125$ is about 2 holes/plaquette if the space is fully occupied by these plaquettes. Due to the spatial inhomogeneity, in some areas, the doping level may have reached higher values although the nominal doping level is at $p = 0.10$ in present UD10 sample. (2) The coherence-like peaks appear at about $\pm 25\,\text{mV}$ (Fig. 5C), these precursor gaps are quite close to that measured in Bi2201 superconducting samples (*36*). However, we must emphasize that, we call this gap at about 25 mV as the precursor one for the forthcoming superconductivity, it does not mean this is just the superconducting gap since the superfluid density is still quite low. In the samples with more doping, when superconductivity appears, it is the similar weight of DOS that piles up around the Fermi energy and yields a pair of coherence peaks with more shaper feature at lower energies (*23,36*). Our results strongly suggest that Cooper pairing may be established through the coherent charge freedom of two doped holes mainly on oxygen sites via strong entanglement and interaction of the magnetic superexchange within the $4a_0 \times 4a_0$ plaquette.

Concerning local pairing in cuprates, there are plenty supportive information. The large Hubbard U in the system may lead to the quantum spin liquid state with highly fluctuating spin singlets, this is the so-called RVB (*1*) model. The original RVB model postulates that these spin singlets are just Cooper pairs, when they can move, a superconducting state can be achieved by establishing the long range phase coherence. However, the parent phase was later shown to be an AF Mott insulator, not the RVB state. By doping holes, this AF order is destroyed and electric conduction becomes better. It has long been perceived that the superconductivity in cuprates arises from diluted superfluid density on top of a strong pairing background (*37-39*), and this is consistent with strong superconducting fluctuations in underdoped region (*40,41*). The electronic phase separation can be induced by the frustrated quantum interactions in extremely underdoped cuprates (*7*), and then the doped holes may appear in the conducting stripes or domain walls of the sample (*7,30,31*). These doped holes may be locally paired in the conducting stripes, forming the upcoming superconductivity (*8,30,31*). Besides, another assembled picture would be that a local pairing is established through doped holes on the background with strong magnetic superexchange effect. This is quite similar to the situation depicted by Emery, he argued that the charge carriers may be the holes doped to the O(2*p*) states and the pairing is mediated by strong coupling to local spin configurations on the Cu sites (*33*). In an alternative picture, the pairing attraction between two holes was also illustrated by the quantum phase strings created by the hopping of holes, but not by the long-range antiferromagnetic nor the RVB pairing in the spin background (*42*). Furthermore, pictures concerning a Bose metal phase with composite bosons intervening between superfluid and insulator were proposed to interpret the transport properties and Fermi arc behavior in pseudogap region (*43,44*). Although these models show qualitative consistency with our data, it remains however curious to know how to interpret the unique pattern of nematic structure within a plaquette observed in our experiment. Actually, a recent theoretical approach finds that unconventional superconductivity is spectacularly enhanced if the hopping matrix elements (corresponding to the coherent weight here) are periodically modulated in a stripe-like pattern, which shows further consistency with our experiments (*45*). Combining all these facts and theoretical considerations, we conclude that the local Cooper pairing may exhibit as a confined nematic state within a $4a_0 \times 4a_0$ plaquette, which is induced by the quantum entanglement of two holes on the background of strong magnetic superexchange effect. Our experiments shed new light to unravel the puzzle of Cooper pairing in cuprate superconductors, and should significantly stimulate theoretical considerations about the basic pairing function.

**Acknowledgments:** We thank H. Q. Luo for the efforts in growing the single crystals, and we also acknowledge helpful discussions with Z.-Y. Weng, J. Schmallian, and I. Eremin. **Funding:** This work was supported by National Natural Science Foundation of China (12061131001, 11974171, 92065109), and Strategic Priority Research Program of Chinese Academy of Sciences (XDB25000000). **Author contributions:** STM/STS measurements and analysis were performed by H. Z. L., H. L., Z. H. W., S. Y. W., H.Y. and H.-H.W. H. L. and Z. H. W. measured transport properties of samples. H.-H.W., H.Y. and H. Z. L. wrote the paper. H.-H. W. coordinated the whole work. All authors have discussed the results and the interpretations. **Competing interests:** The authors declare that they have no competing interests. **Data and materials availability:** All data needed to evaluate the conclusions in the paper are present in the paper and/or the Supplementary Materials. Additional data related to this paper may be requested from the authors.

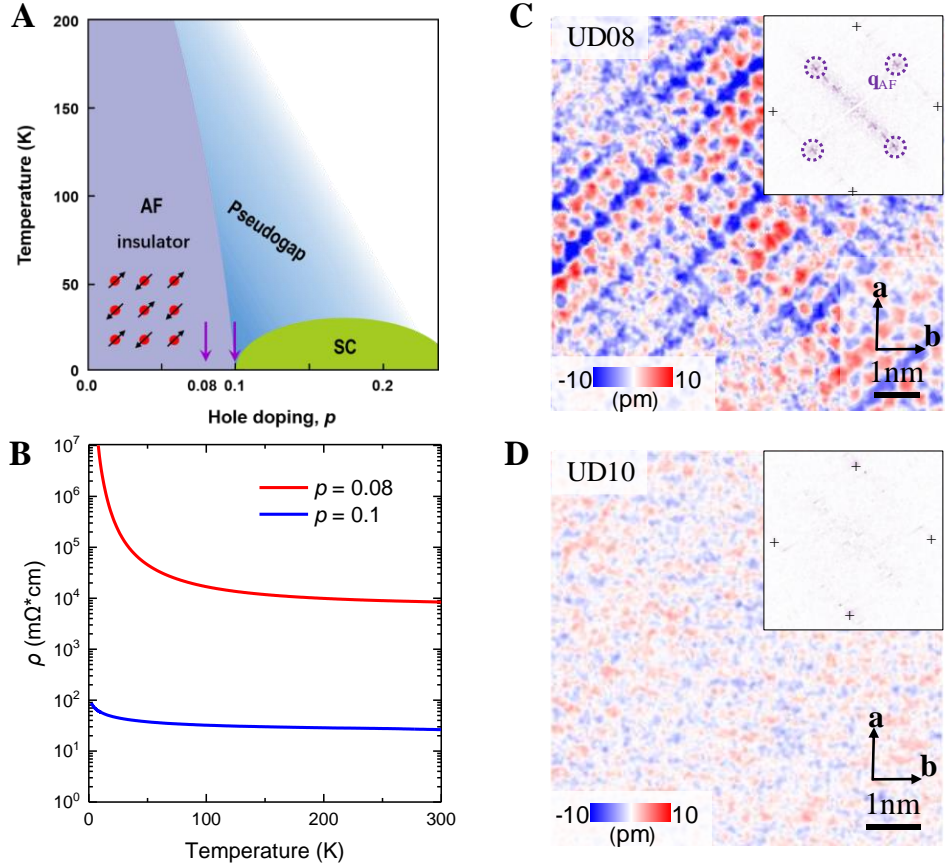

**Fig. 1. Insulating behavior and AF order in extremely underdoped La-Bi2201 samples.** (**A**) Phase diagram of La-Bi2201 systems. The arrows indicate the doping levels of two samples studied here. (**B**) Temperature dependent resistivity measured in two extremely underdoped La-Bi2201 samples. (**C** and **D**) Spin-difference topographies derived by the subtraction of the topographic images measured by the Cr tip under magnetic fields with opposite directions in UD08 and UD10, respectively. The inset in C (or D) shows the FT pattern of C (or D), and the dashed circles in C indicate the pattern of the AF order observed in UD08.

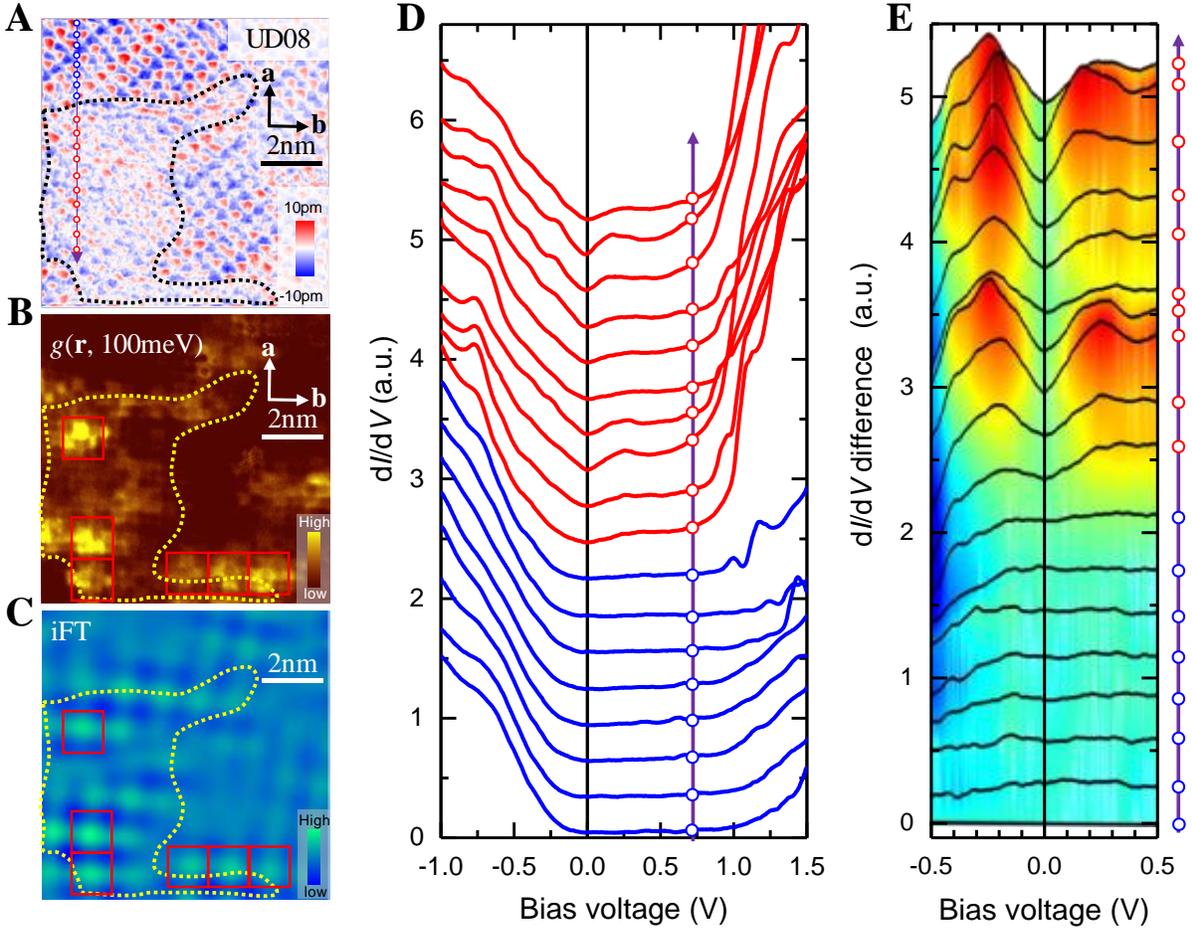

**Fig. 2. Spatial variation of AF order and low-energy DOS in UD08.** (**A**) Spin-difference topography (Setpoint condition: $V_{set}$ = −400 mV, and $I_{set}$ = 50 pA). The AF order is weak in the region surrounded by the dashed line. (**B**) Differential conductance mapping measured in the same region of A (V = −400 mV, I = 100 pA) at 0 T. The red squares indicate the obvious patterns of $4a_0 \times 4a_0$ appearing in the region with extremely weak AF order. (**C**) Inverse Fourier transform (iFT) image shown in the same area of B, and it is calculated from the characteristic spots corresponding to $4a_0 \times 4a_0$ periodic patterns in the FT image (fig. S2B). (**D**) A set of tunneling spectra measured along the purple line in A ($V_{set}$ = −1 V, and $I_{set}$ = 100 pA). The spectra are offset for clarity. (**E**) Difference spectra of the tunneling spectra in C by subtracting the first spectrum as the background. They are presented in a narrow bias range in order to show the detailed structure at low energies near $E_F$. All the data are taken by using an SP Cr tip.

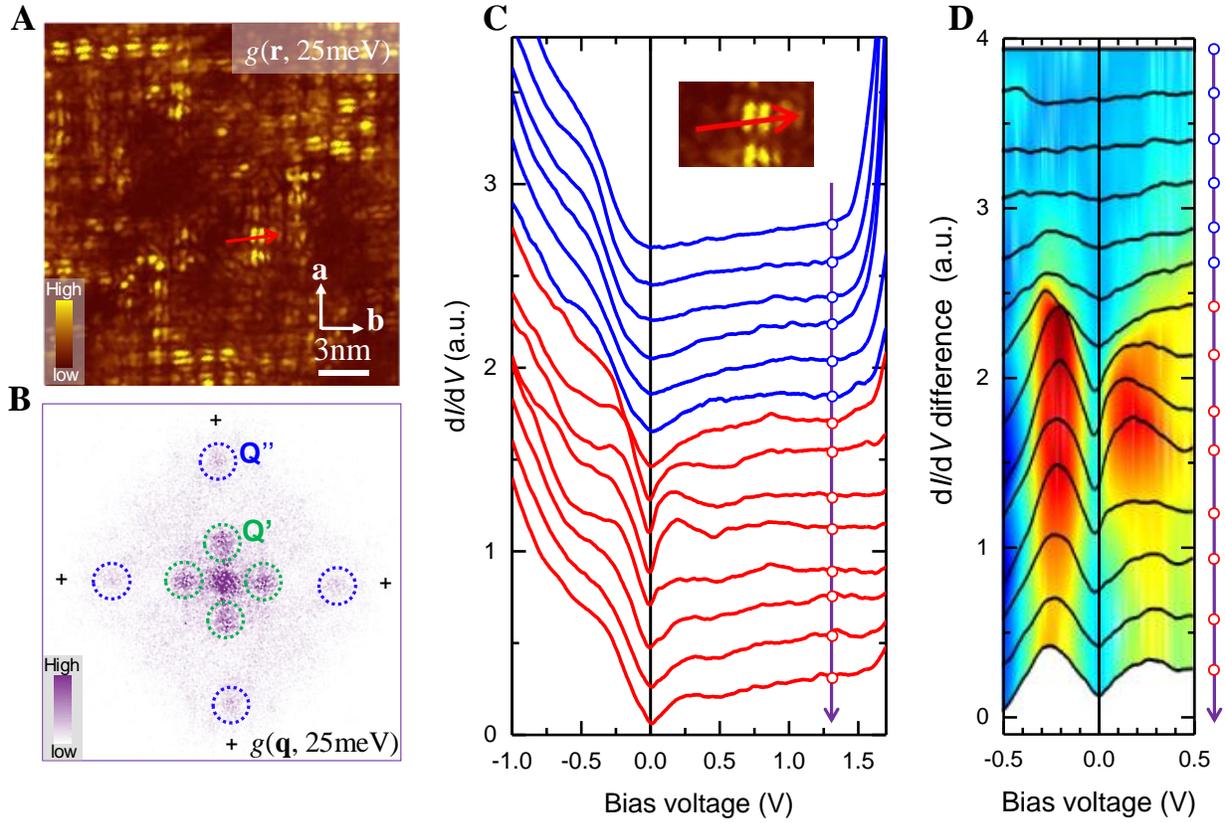

**Fig. 3. Electronic modulations and the related evolution of DOS in UD10.** (**A**) Differential conductance mapping measured at 25 mV ($V_{set} = -250$ mV, and $I_{set} = 100$ pA). (**B**) The FT pattern of the d$I$/d$V$ mapping shown in A. **Q'** spots correspond to the checkerboard patterns with the period of $4a_0$ in the real space, while **Q"** spots correspond to the nematic modulation with the period of $4a_0/3$ in the real space. (**C**) Spatially resolved tunneling spectra measured along the arrowed line in A at $V_{set} = -1$ V, and $I_{set} = 100$ pA. The spectra are offset for clarity. (**D**) Difference spectra of the tunneling spectra in C by subtracting the first spectrum as the background. They are presented in a narrow bias range near $E_F$. All the data are taken by using a Cr tip.

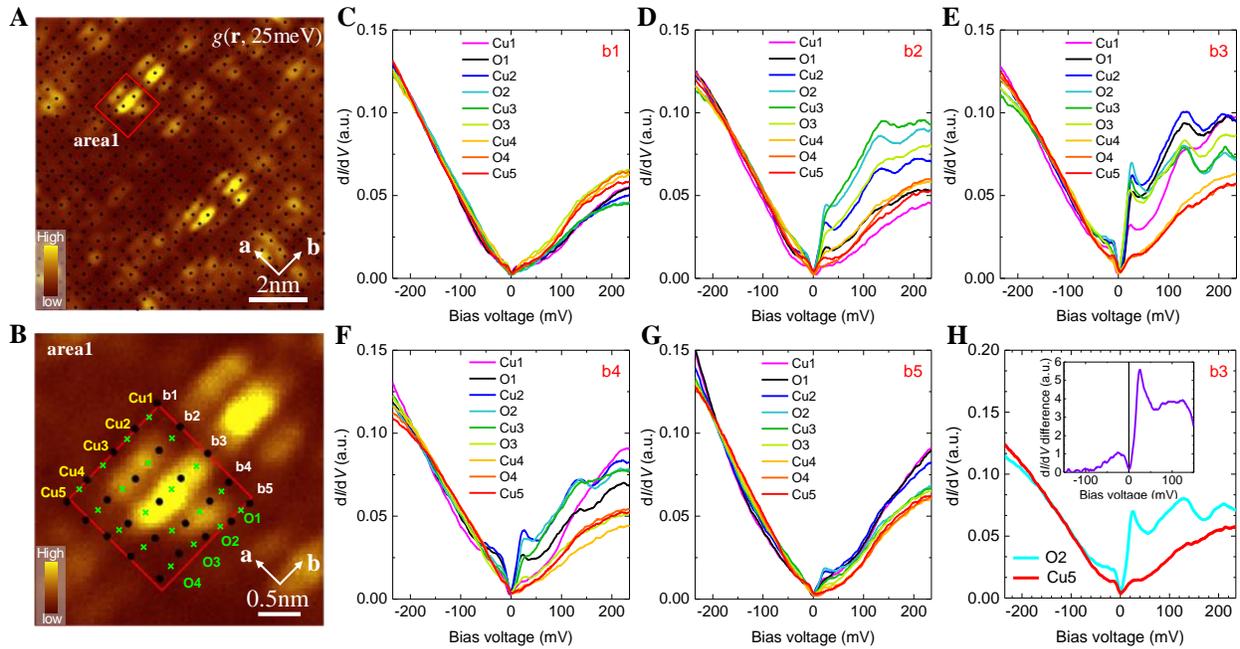

**Fig. 4. Detailed electronic structure within a $4a_0 \times 4a_0$ plaquette in UD10.** (**A**) An enlarged view of the d$I$/d$V$ mapping measured at 25 mV ($V_{set}$ = −250 mV, and $I_{set}$ = 100 pA). Black dots represent the positions of Bi atoms obtained in the topography taken simultaneously with the d$I$/d$V$ mapping, and the lattice is distorted due to the supermodulation on the top surface. Cu atoms underneath are supposed to be at the same positions as Bi atoms on the surface. The red square indicates a typical plaquette of $4a_0 \times 4a_0$. (**B**) A detailed structure of the $4a_0 \times 4a_0$ plaquette highlighted in A. Three nematic modulation structure can be seen as bright bars. The positions of some oxygen atoms (marked by crosses) are determined as the middle points of the neighboring Cu atoms roughly along the *b*-axis. (**C** to **G**) Tunneling spectra measured at the positons of Cu/O atoms along five Cu-O-Cu chains (b1-b5) roughly in the direction of the *b*-axis within the typical plaquette. (**H**) Tunneling spectra measured at O2 and Cu5 positions in b3 chain. The O2 positon in b3 chain is in the brightest bar, and the Cu5 position is away from three bright bars. The inset shows the difference spectrum of these two tunneling spectra. All the data are taken by using a tungsten tip.

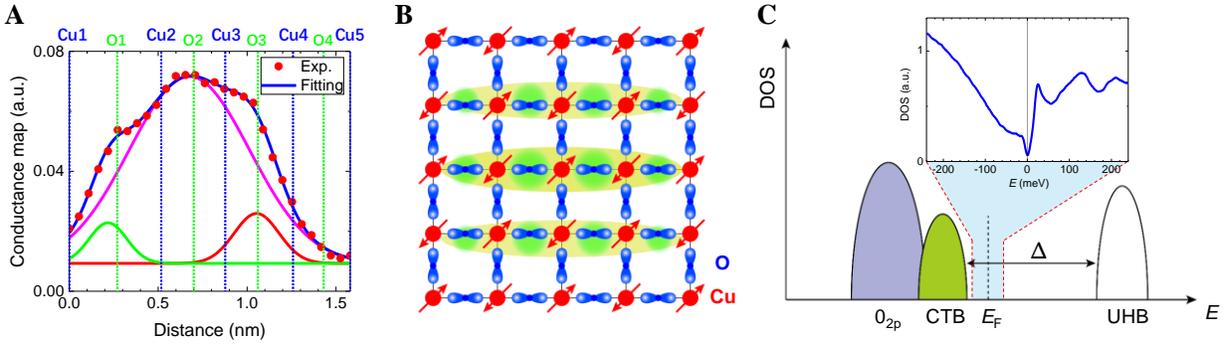

**Fig. 5. Determination and theoretical model of electronic states on atomic scale in UD10.** (**A**) Determination of the local DOS distribution on atomic sites by the multi-peak Gaussian fitting. The differential conductance along the b3 Cu-O-Cu chain (red filled circles taken from Fig. 4B) can be well fitted by the blue solid line constructed by three assembled Gaussian functions (red, pink and green solid lines) with peak positions locating at oxygen sites (blue solid lines). (**B**) Cartoon picture showing three bars of enhanced DOS with nematic modulations as highlighted by yellow ellipsoids in atomic scale. The enhanced DOS at about ±25 meV around the central O2 and O3 atoms are highlighted by the yellow-green patches. (**C**) Schematic picture of the DOS induced by the chemical doping emerging near $E_F$ within the charge transfer gap. The inset shows the spectrum measured on the O2 atom revealing a coherence-like peak of DOS at about ±25 mV.

# Supplementary Materials for

**Real-space pairing through a confined local nematic state in cuprate superconductors**

Huazhou Li[1†], Han Li[1†], Zhaohui Wang[1], Siyuan Wan[1], Huan Yang[1*], Hai-Hu Wen[1*]

* Corresponding authors: huanyang@nju.edu.cn (H. Y.); hhwen@nju.edu.cn (H.-H.W.)

**Materials and Methods**

The single crystals of $Bi_2Sr_{2-x}La_xCuO_{6+\delta}$ were grown by using the traveling-solvent floating-zone technique (*15*). The samples look very shinny with a very good crystallinity checked by x-ray diffraction pattern. STM/STS measurements were carried out in a scanning tunneling microscope (USM-1300, Unisoku Co., Ltd.) with the ultrahigh vacuum up to $10^{-10}$ torr. During the STM experiment, we used two kinds of electrochemically etched tips, i.e., the spin-polarized chromium tip (*46*) and the conventional tungsten tip. The Cr tip was electrochemically etched by 3 mol/L NaOH solution with the immersed end covered by a polytetrafluoroethylene tube with a certain length (*47*). After the etching, the Cr tip was transferred into the STM chamber, and the spin polarization feature was characterized on the cleaved surface of single crystalline $Fe_{1+x}Te$ (*13,14,48*). The magnetic field is applied perpendicular to the surface (*ab* plane) of the sample during the measurements. The STM data are taken at about 2.0 K by using the chromium tip, and those are taken at about 1.6 K by using the tungsten tip. A typical lock-in technique is used with an ac modulation of 987.5 Hz.

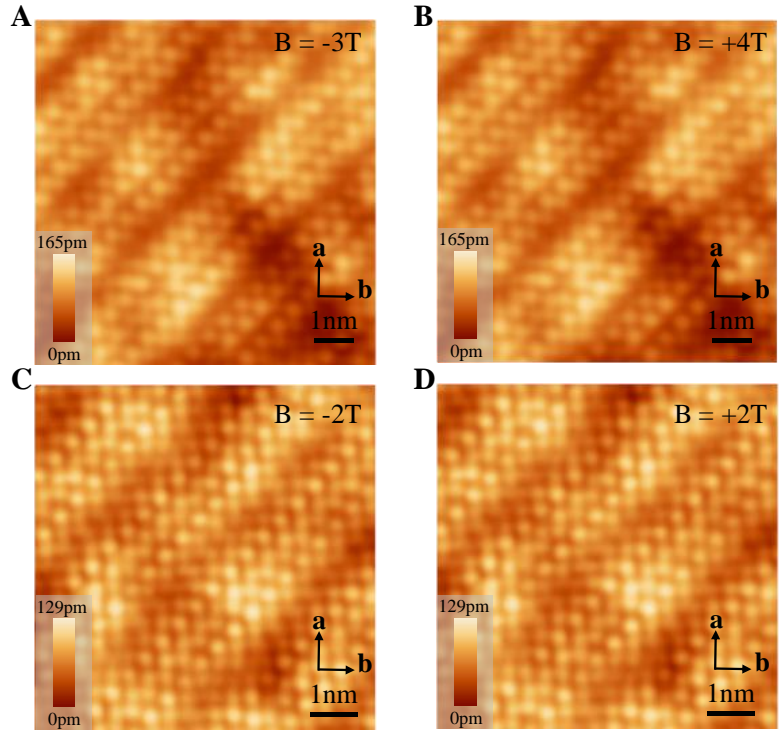

**Fig. S1. Spin resolved topographies measured at magnetic fields with opposite directions.** (**A** and **B**) Topographic images acquired by using the SP-Cr tip in UD08 and at a magnetic field of −3 and +4 T, respectively ($V_{set} = -400$ mV, and $I_{set} = 100$ pA). (**C** and **D**) Topographic images acquired by using the SP-Cr tip in UD10 and at a magnetic field of −2 and +2 T, respectively ($V_{set} = -400$ mV, and $I_{set} = 50$ pA). Spin-difference topography shown in Fig. 1C (or in Fig. 1D) is derived by the subtraction of B from A (or C from D).

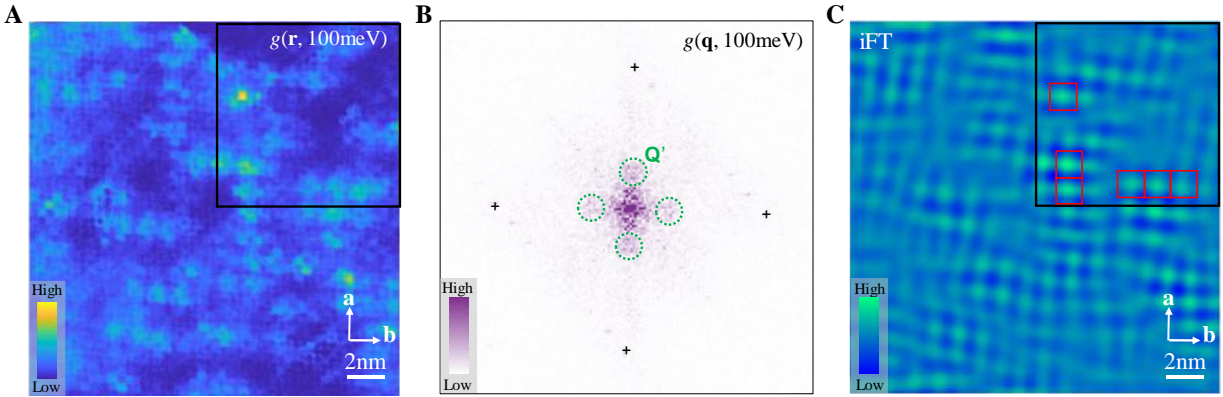

**Fig. S2. Differential conductance mapping measured in a large area in UD08 by using a Cr tip.** (**A**) The d$I$/d$V$ mapping measured at 100 mV ($V_{set}$ = −400 mV, and $I_{set}$ = 100 pA). The area framed in the black square is that shown in Fig. 2B. (**B**) FT pattern to the d$I$/d$V$ mapping of A. The spots of Q' correspond to the checkerboard patterns with the period of $4a_0 \times 4a_0$ in the real space. (**C**) Inverse FT (iFT) pattern to the Q' spots within the areas in circles of B. One can see clear $4a_0 \times 4a_0$ checkerboard patterns, and the patterns highlighted by red squares in Fig. 2B are also indicated by red squares in C.

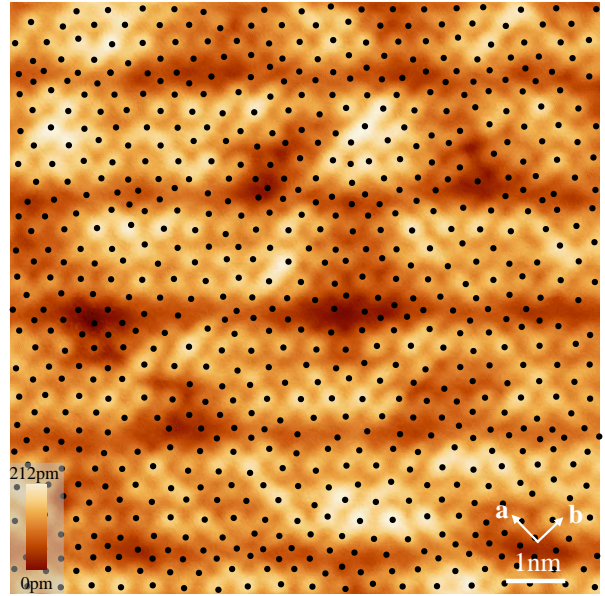

**Fig. S3. Topography measured simultaneously with the d$I$/d$V$ mapping shown in Fig. 4A by using a tungsten tip.** Setpoint: $V_{set}$ = −250 mV, $I_{set}$ = 100 pA. The black dots are the positions of Bi atoms on the surface, and they are supposed to be the positions of Cu atoms underneath.

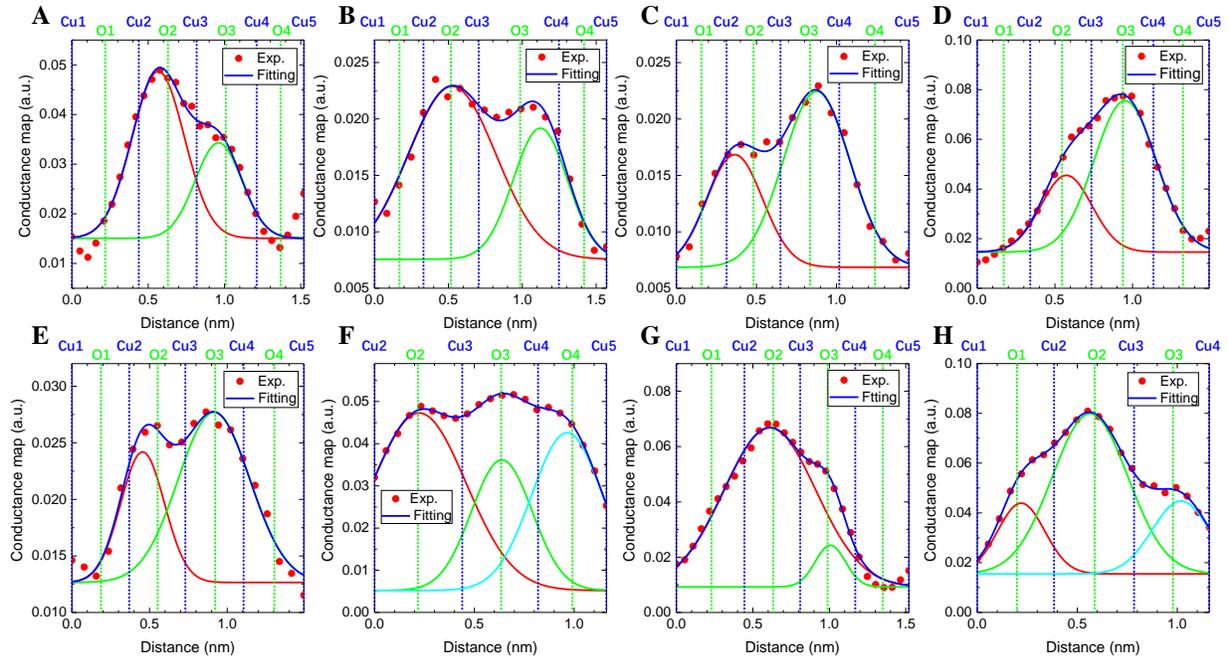

**Fig. S4. Determination of the local DOS distribution on atomic sites by the multi-peak Gaussian fitting.** The solid symbols are the data of the spatial dependent differential conductance along the Cu-O-Cu chains of b3 across several bright central bars of confined nematic electronic state in checkerboard patterns in UD10. The experimental data can be well fitted by Gaussian functions with peak positions locating at oxygen sites. The two major peaks all locate near the O2 and O3 potions.

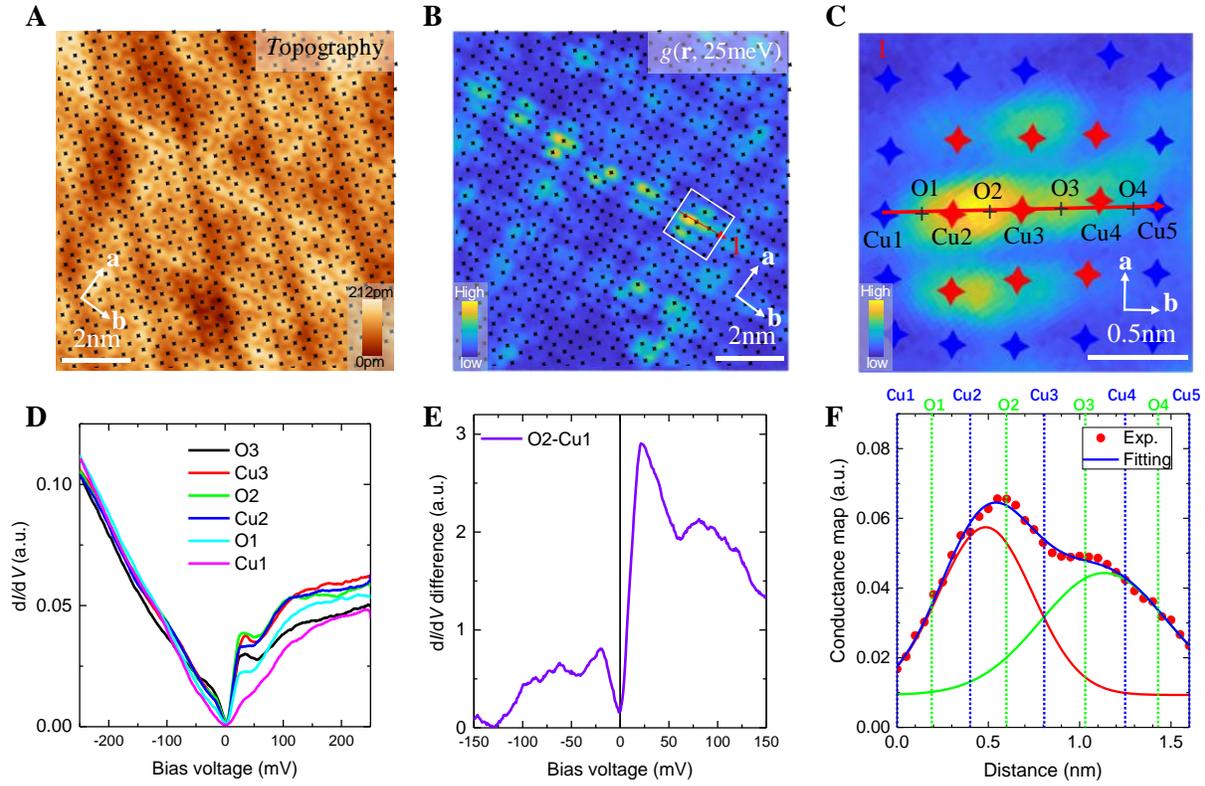

**Fig. S5. Electronic modulations and the related evolution of DOS in UD10 measured by an SP Cr tip.** (**A** and **B**) Topography and d$I$/d$V$ mapping measured simultaneously in the same area ($V_{set}$ = −250 mV, and $I_{set}$ = 100 pA). The black dots in A are the positions of Bi atoms on the surface or the Cu atoms underneath. (**C**) A detailed structure of the $4a_0 \times 4a_0$ modulation highlighted in B. Three nematic modulation structure can be seen as bright bars. (**D**) Tunneling spectra measured at the positons of Cu/O atoms along the b3 Cu-O-Cu chain (red arrowed line) in C. (**E**) Difference spectrum of two tunneling spectra measured at O2 and Cu1 positions in b3 chain. (**F**) Differential conductance along b3 in C, and it can be well fitted by Gaussian functions with peak positions locating at oxygen sites. The two major peaks locate at O2 and O3 potions.

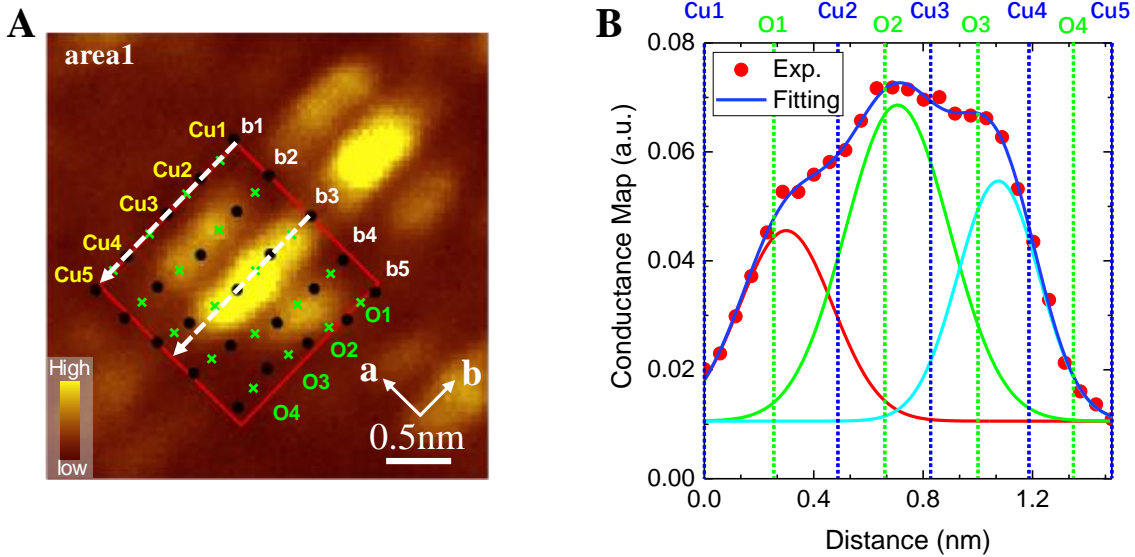

**Fig. S6. Determination of the local DOS distribution along a straight line by the multi-peak Gaussian fitting.** (**A**) Differential conductance mapping the same as Fig. 4B. Since the Bi atoms along the central bright bar deviate from the straight line at Cu4/Bi4 and Cu5/Bi5 sites along the b3 chain, we plot a straight line (the dashed arrowed line b3) parallel the Cu-O-Cu chain of b1 and starting from the Cu1/Bi1 site. (**B**) Solid symbols are the data of the spatial dependent differential conductance along the straight line of b3 shown in A. The experimental data can be well fitted by Gaussian functions with peak positions also locating at oxygen sites. The two major peaks also locate near the O2 and O3 potions.